\documentclass[twocolumn,preprintnumbers,amsmath,amssymb]{revtex4}

\usepackage{graphicx,color}
\usepackage{dcolumn}
\usepackage{bm}
\usepackage{CJK}

\begin{document}

\title{Third minima in actinides - do they exist?}

\author{M. Kowal }
\email{m.kowal@fuw.edu.pl}
\author{J. Skalski}
\affiliation{National Centre for Nuclear Research, Ho\.za 69,
PL-00-681 Warsaw, Poland}

\date{\today}

\begin{abstract}

 We study the existence of third, hyperdeformed minima in a number of
  even-even Th, U and Pu nuclei using the Woods-Saxon
  microscopic-macroscopic model that very well reproduces first and
  second minima and fission barriers in actinides.
  Deep ($3 \div 4$ MeV) minima found previously by \'Cwiok et al. are found
  spurious after sufficiently general shapes are included.
  Shallow third wells may exist in $^{230,232}$Th, with
  IIIrd barriers $\le$ 200 and 330 keV (respectively).
  Thus, a problem of qualitative discrepancy
 between microscopic-macroscopic and selfconsistent predictions is
  resolved. Now, an understanding of experimental results on the
  apparent third minima in uranium becomes an issue.
\end{abstract}

\maketitle

 The topic of third minima in actinides, supposedly more deformed than
  the superdeformed (SD) ones, and for this reason sometimes called
  "hyperdeformed", is surrounded by some uncertainty. Their existence
   is inferred from an analysis of the observed transmission resonances
  in the prompt fission probability in (n,f), (d,pf), ($\gamma$,f) and
  (e,f) reactions. Using one-dimensional models for tunneling
  one tries to fit the data by considering various possiblities of
  rotational or vibrational states, either in the second or the third well,
  and choosing unknown barrier parameters. The best fit gives
   energy at the third barrier and minimum, and possibly the moment of
   inertia in the IIIrd well.
   Additional check of the hypothesis may be obtained by measuring
    fission fragment angular distribution.
   First interpretation of the third minima in thorium \cite{Blons1} was
   supported by later studies \cite{Blons2,Blons3,Blons4,ph,e}
  in $^{230,231,232,233}$Th which found them rather shallow $\approx$300 keV.
  Recent experiments of the Debrecen-Munich group
 report third minima in a number of nuclei ($^{232,234,236}$U,$^{232}$Pa)
 \cite{HyperU,232Pa,Hyper232U,Thirolf}. Based  on resonance energies inbetween
  the first and the (higher) second barrier and their small widths,
   they infer a third barrier at least as high as the second one and
  deep third minima.
   There are not any direct measurements of the third minima so
     some reservations concerning the interpretation would be in order.

 Theoretical predictions of the third minima in actinides present
 somewhat confusing picture. While the presence of the second minima at
  deformation close to $\beta_{2}=0.6$ is a common feature of both
 macroscopic-microscopic (since Strutinsky \cite{St67}) and self-consistent calculations,
 the third minima appear or disappear from calculation to
 calculation. The first calculated IIIrd minima \cite{MN,HM} were predicted
  shallow, 0.5 to 1.5 MeV, with the third barrier higher than the
  second one in thorium. Bengtsson et al. \cite{Bengt} had very shallow minima
 (0.5 MeV or less), with the lower third barrier.
  Qualitatively different results, with prominent 3-4 MeV deep, and often
  multiple, third minima, also in U and heavier nuclei,
 were presented by S.~\'Cwiok et al. \cite{Stefan,Stefan1,Stefan2}.
  These calculations used a richer set of axially symmetric deformations
  than the previous ones, up to the multipolarity $\lambda=7$.
 Results of the Hartree-Fock calculations indicate that presence or absence
of third minima depends very much on the effective interaction.
However,
  minima in U and Pu, if they appear at all, are rather shallow and nearly
  degenerate with, or even lower than, the ground states \cite{Bender}.
  Therefore, they do not fit Debrecen-Munich results.
 On the other hand, calculations with the Gogny force \cite{Gogny} and SkM$^{*}$ \cite{HF+BCS} (Hartree-Fock plus BCS) predict
 shallow minima in some Th isotopes, compatible with Th data \cite{Blons4},
  while the other work \cite{Delar} provides shallow minima at the top of the
  first barrier, not of the type suggested by the experimental papers.

  The aim of this work was to check the existence of the third minima in
  actinides within the Woods-Saxon microscopic-macroscopic model used in
 \cite{Stefan,Stefan1}
  by extending the number of included deformations. An impulse to this came
  from first repeating old calculations and realizing that  they produce
  IIIrd minima also in many heavier actinides, in which
  they were not found experimentally. Moreover, the calculated
  minima with larger octupole deformations have
   quadrupole moments (calculated assuming the sharp-surface distribution
  and the radius parameter 1.16 fm) $Q\sim 170$ b, disturbingly close to
  the scission region. The contemplated possibility was that
  some IIIrd minima in \cite{Stefan,Stefan1} are just intermediate configurations
  on the scission path, whose energy was calculated erroneously because
  of limitations of the admitted class of shapes.
{\it Method} Third minima were searched by exploring energy
 surfaces in a region
 of deformations beyond the second minima. The energy is calculated within
 the microscopic-macroscopic method with a deformed Woods-Saxon potential.
 Nuclear shapes are defined in terms of the nuclear surface
  \cite{WS}
   \begin{eqnarray}
  \label{shape}
    R(\theta,\varphi)&=& c(\{\beta\})
    R_0 (1+\sum_{\lambda>1}\beta_{\lambda}
   Y_{\lambda 0}(\theta,\varphi)+
    \nonumber \\
   &&
    \sum_{\lambda>1, \mu>0, even}
   \beta_{\lambda \mu} Y_{\lambda \mu c} (\theta,\varphi) +
   \nonumber \\
   &&
    \sum_{\lambda>1, \mu>0, odd}
   \beta_{\lambda \mu} Y_{\lambda \mu s} (\theta,\varphi))  ,
 \end{eqnarray}
  where $c(\{\beta\})$ is the volume fixing factor, the real-valued spherical
  harmonics $Y_{\lambda \mu c}$, with $\mu>0$
  even and $Y_{\lambda \mu s}$, with $\mu>0$ odd, are defined in terms of the
  usual ones as: $Y_{\lambda \mu c}=(Y_{\lambda \mu}+Y_{\lambda -\mu})/
 \sqrt{2}$ and $Y_{\lambda \mu s}=-i(Y_{\lambda \mu}+Y_{\lambda -\mu})/
 \sqrt{2}$.
  Such shapes have at least one symmetry plane $y-z$.
  Although nonaxiality modifies energy landscapes in the region of
  the third barrier, the main effect comes from a proper treatment
  of the axially symmetric shapes. Therefore, we explore here the first
  part of the Eq. (1).

  Two models were used for the macroscopic part: the Yukawa plus exponential
  \cite{KN}, used by us up to now in many studies, and recently given version
 of the Liquid Drop \cite{LSD} with the mean-curvature term included in
 the deformation-dependent part of energy.
  The reason for using the latter was to check  whether
  double third minima in \cite{Stefan,Stefan1,Stefan2} are not related
  to some singularity of the used Yukawa-plus-exponential macroscopic energy.

   The s.p. potential parameters used in the present work, as well as the way
  the shell- and pairing corrections are the same as in
  a number of our previous studies, e.g. in \cite{KOW10,kowskal,IIbarriers}.
  Let us emphasize that the used model very well reproduces first \cite{KOW10},
  and second barriers \cite{IIbarriers} and second minima \cite{kowskal}
  in actinide nuclei.
{\it Removig shape limitation}
  A general difficulty in studying very deformed nuclear shapes lies in their
  effective parametrization. The Eq. (\ref{shape}), natural for small and
  moderate deformations, becomes problematic for this purpose.
  Rather large (more than six) number of deformations is necessary for
  a sufficient accuracy. Their sizes may be misleading,
   for example, there are sets $\{\beta_{\lambda }\}$ corresponding to
  {\it huge} elongations with {\it quite moderate} $\beta_{2}$.
  Such is the case of two kinds of IIIrd minima in actinides
  shown in \cite{Stefan,Stefan1,Stefan2}. With similar $\beta_2\approx 0.85$
 they might
  be thought close in elongation. But they are not: they correspond to
  different quadrupole moments $Q\approx 120$ and 170 b.
  The energy mapping in many-dimensional space becomes a problem.
  A reduction of dimension via the minimization over some of the deformations
   often leads to an energy surface composed from disconnected patches,
  corresponding to multiple minima in the
  auxiliary (those minimized over) dimensions.
  If a smooth look of such a surface hides the jumps between minima, it
  suggests a false picture of the barrier.

  It may be for these difficulties that an important shape limitation
  of Eq. (\ref{shape}) did not attract much attention. The dipole deformation
 $\beta_1$ is omitted there, as corresponding to a shift of the origin
  of coordinates which leaves energy (always calculated in the center of mass frame) invariant.
   However, this is true only for {\it weakly} deformed shapes. For large
   elongations, $\beta_1$ acquires a meaning of a real shape variable.

   The parametrization (\ref{shape}) is not unique, as the same
   shape has many expansions corresponding to various shifts of the
   origin of coordinates. Hence, one could think that $\beta_1$ may
   be always transformed out. However, the shift invariance is lost
   beacause a) the expansion (\ref{shape}) is always truncated,  b) for well
   constricted shapes, large shifts lead outside the formula (\ref{shape})
    making the radius multivalued.
   In practice, admitting nonvanishing $\beta_1$ in (\ref{shape}),
  \begin{equation}
   \label{shape1}
    R(\theta,\varphi)= c(\{\beta\})R_0 (1+\sum_{\lambda=1}\beta_{\lambda}
   Y_{\lambda 0}(\theta) )  ,
  \end{equation}
   turns out to be an effective way of including an important class of
   smoothly constricted elongated shapes, which are {\it unreachable} when
   using a reasonably truncated expansion (\ref{shape}) {\it without} it.
  One should notice that the parametrization (\ref{shape1}) is redundant:
  the remaining invariance to small shifts of the origin of coordinates
   means that some shapes have many parametrizations with different $\beta_1$.

{\it Disappearence of spurious minima}
  We considered even-even nuclei $^{230,232}$Th, $^{232,234,236,238}$U and
  $^{240}$Pu. Axially symmetric shapes were assumed. Two sets of energy
  calculations were done: A) maps $(\beta_2,\beta_3)$ with minimization
  over $\beta_4$-$\beta_8$ (7D) and B) in a hypercube $\beta_1$-$\beta_6$ (6D)
 from which various 2D maps could be generated. The grid points were
 (with step in all directions $=0.05$): $-0.35$-0 in $\beta_1$, 0.55-1.5 in $\beta_2$, 0-0.35 in
  $\beta_3$, $-0.10$-0.35 in $\beta_4$, $-0.2$-0.2 in $\beta_5$,
 $-0.15$-0.15 in $\beta_6$.
  To improve limits on barrier heights
  we calculated energies along chosen continuous paths in 8D space
   $\beta_1$-$\beta_8$.

  Results of A) essentially reproduce landscapes in \cite{Stefan,Stefan1}.
  Beyond the axially- and reflection-symmetric second minima, the
   reflection-asymmetry lowers the second saddles.
   Even more (shell correction of $\approx$10 MeV) it lowers energy for still
  larger elongations.
\begin{figure}[t]
\begin{minipage}[t]{52mm}
\centerline{\includegraphics[scale=0.30,angle=-90]{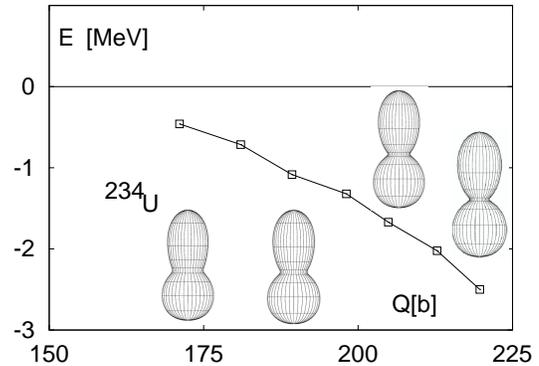}}
\end{minipage}
\caption{{\protect\small Energy along a sequence
  of smoothly elongating shapes parametrized by $\beta_1-\beta_8$, beginning at
   the apparent third minimum with larger octupole deformation
   $\beta_3\approx 0.6$ in $^{234}$U. Quadrupole moments calculated for
   the sharp-surface distribution and the radius parameter $r_0=1.16$ fm.
 Depicted shapes correspond to: $Q=171$ (the apparent IIIrd minimum),
   189, 205 and 220 b.
   }}
\end{figure}
   Double IIIrd minima with $\beta_3\approx 0.3$ and 0.6 are obtained with
   the Yukawa-plus-exponential macroscopic energy; the
  less mass-asymmetric ones are deeper in Th and higher in U and Pu.
 With the LSD macroscopic energy,
 formula \cite{LSD} minima with larger $\beta_3$ (quadrupole moments) vanish
 and the IIIrd minima become single. In both versions the
  third barriers are quite high, $\ge 2.5$ MeV above the IIIrd
  minimum, with the highest ones in thorium.

   The landscape modification obtained by
  including dipole deformation $\beta_1$
  in (\ref{shape}) is decisive for the picture of the IIIrd minima.
   The apparent minima with large $\beta_3$ and quadrupole moments
  disappear completely in the studied nuclei. One can find continuous 8D paths
  starting at the supposed IIIrd minimum and leading to scission,
  along which energy decreases gradually. An example is
  shown in Fig. 1 for $^{234}$U.

   The existence of minima with smaller octupole deformation
  (and quadrupole moments) requires a more detailed study.
  Finding the barrier in many-dimensional space requires
   hypercube calculations.
  The maps $(\beta_2,\beta_3)$ generated by minimization over
 $\beta_4$-$\beta_6$ from the 6D hypercube calculations B) are shown
  in Fig. 2 for $^{232}$U at fixed $\beta_1=$ 0 and $-0.2$.
  The large effect of included new shapes may be seen as a lowering
  of the $\beta_1=0$ third barrier by $\beta_1=-0.2$. Its size of 2-4 MeV
  is common in all studied nuclei, leaving some small barriers in
   $^{230,232}$Th.
   One can note that the heights of the second saddles are practically
  {\it not changed} by $\beta_1$. Hence, in the evaluation
  of secondary fission barriers in actinides (see e.g \cite{IIbarriers})
   this parameter is not necessarily needed.
\begin{figure}[t]
\begin{minipage}[t]{52mm}
\centerline{\includegraphics[scale=0.35, angle=0]{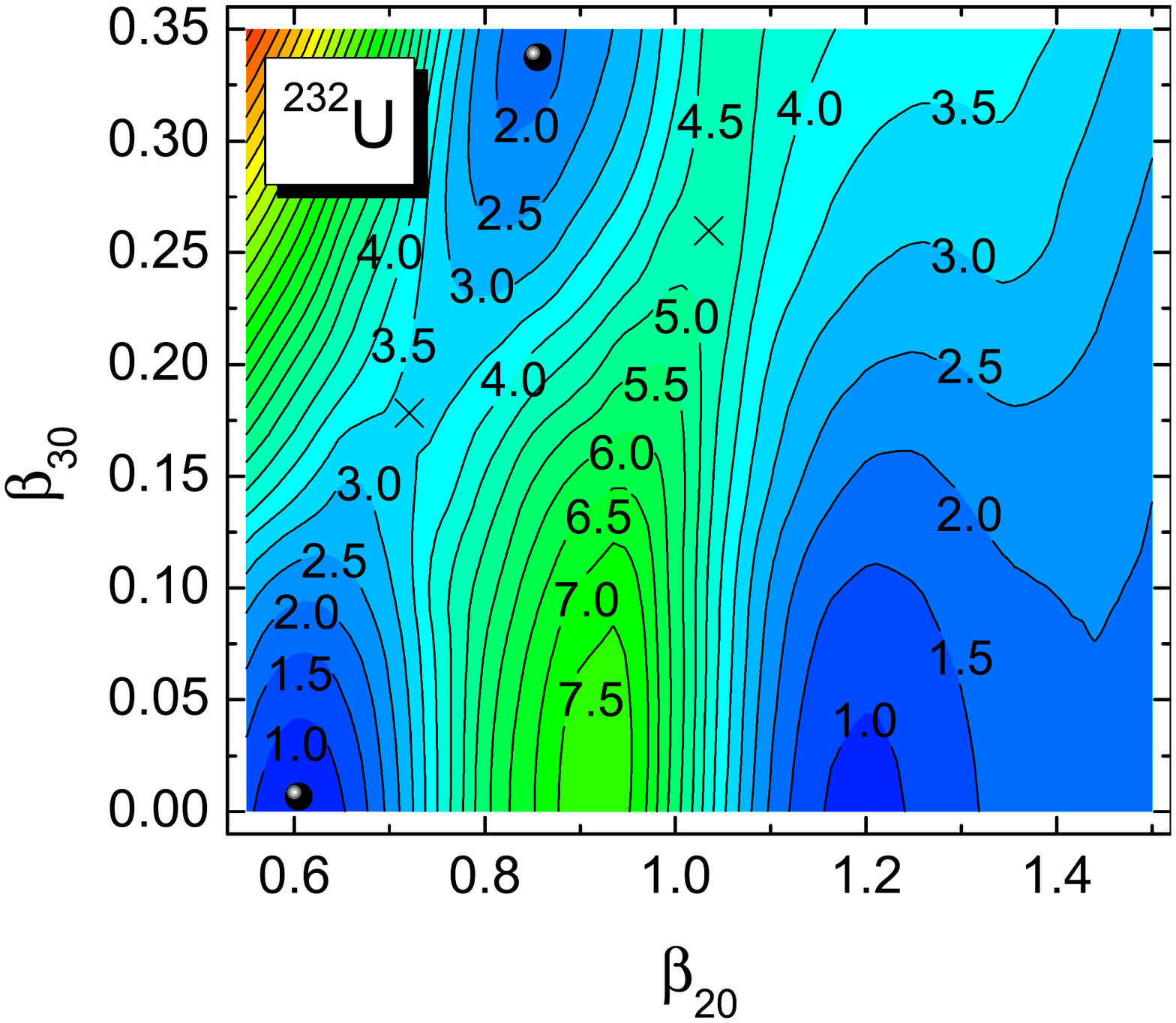}}
\end{minipage}
\end{figure}
\begin{figure}[t]
\begin{minipage}[t]{52mm}
\centerline{\includegraphics[scale=0.35,
angle=0]{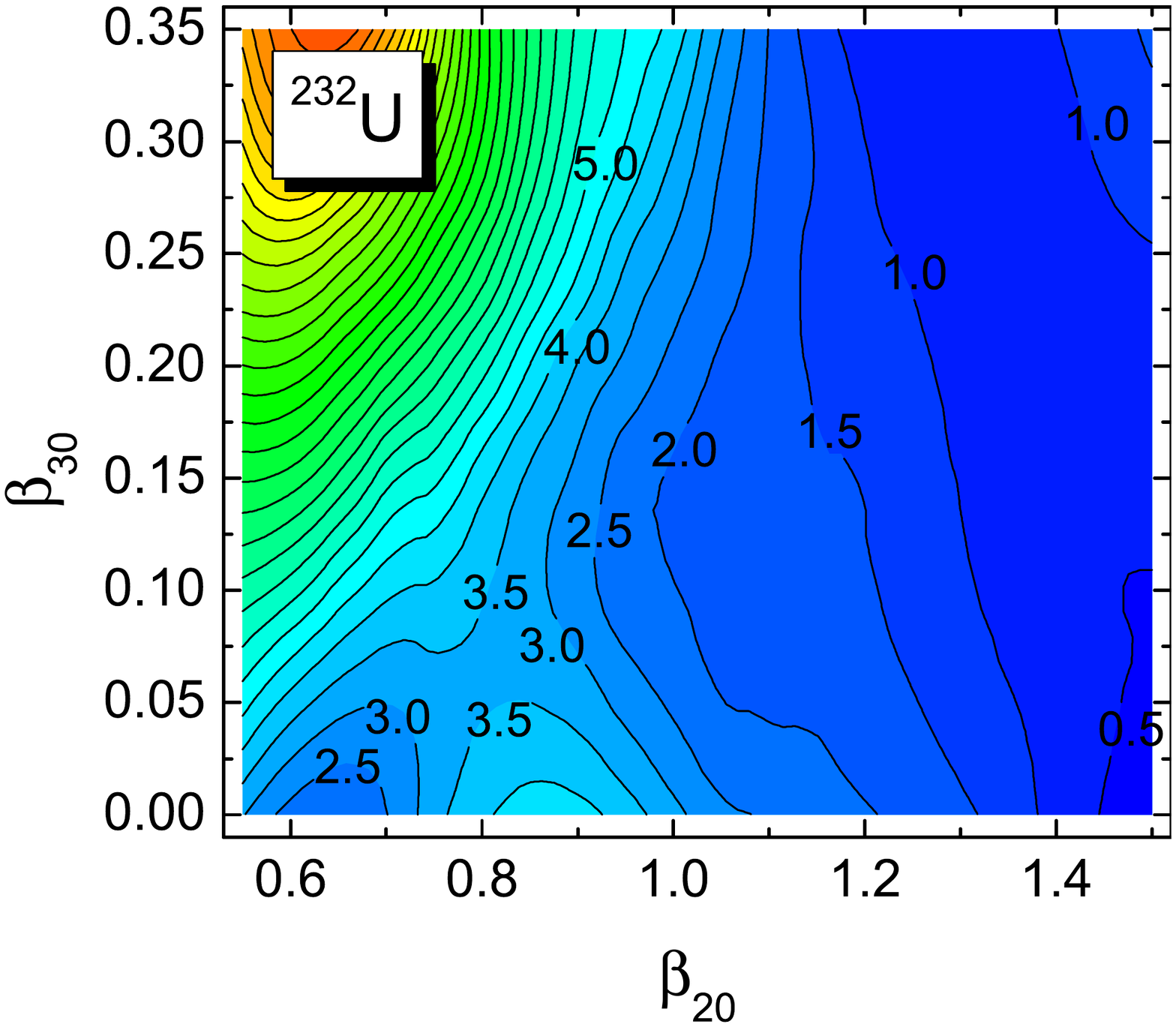}}
\end{minipage}
\end{figure}
\begin{figure}[t]
\begin{minipage}[t]{52mm}
\centerline{\includegraphics[scale=0.35, angle=0]{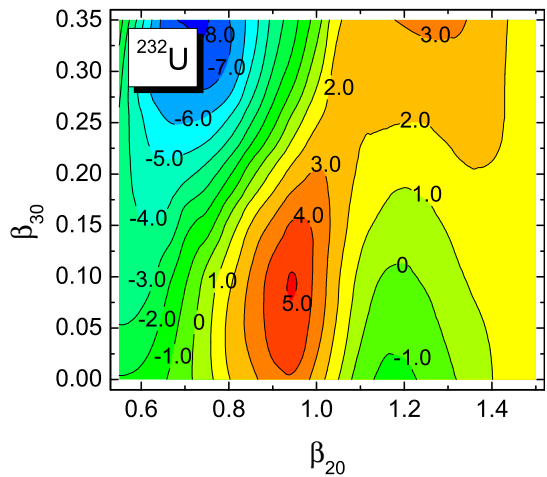}}
\end{minipage}
\caption{{\protect\small Energy maps for $^{232}$U from the 6D
  $\beta_1-\beta_6$ calculation, minimized over $\beta_4,\beta_5$;
  upper panel - $\beta_1=0$, middle panel - $\beta_1=-0.2$. The
  lower panel gives the difference $E_{upper}-E_{middle}$, so the lowering
  of the third barrier by $\beta_1=-0.2$ can be read from it as a positive
   number.
    }}
\end{figure}
  The results B) indicate a large effect of new shapes, but
   are not precise enough for a detailed barrier prediction.
  Our tests show that deformations $\beta_1$-$\beta_8$ are
  needed for fixing minimima up to 100-150 keV, a greater precision
  requires at least $\beta_9$.
  As 8D or 9D hypercube calculations are very time-consuming,
  we find the upper limit on barriers selecting continuous paths in
   8D $\beta_1$-$\beta_8$ space, from
  the supposed IIIrd minima towards scission. Energies vs quadrupole
  moments along such paths are shown in Fig. 3 for $^{232}$Th and
  $^{232}$U. As can be seen, the barrier vanishes in uranium and must be
  smaller than 330 keV in $^{232}$Th. The only other nonzero upper limit
  on the IIIrd barrier of 200 keV we find in $^{230}$Th.
   As by increasing the number of
  deformations one can only lower saddle points, the third barriers could rise
  only by lowering the supposed IIIrd minima. But from our tests
    this effect is of the order of 100 keV,
  so, within the used model, the barrier in $^{232}$Th should not
  be higher than $\approx 450$ keV.
\begin{figure}[r]
\hspace{-4cm}
\begin{minipage}[r]{52mm}
\includegraphics[scale=0.30, angle=-90]{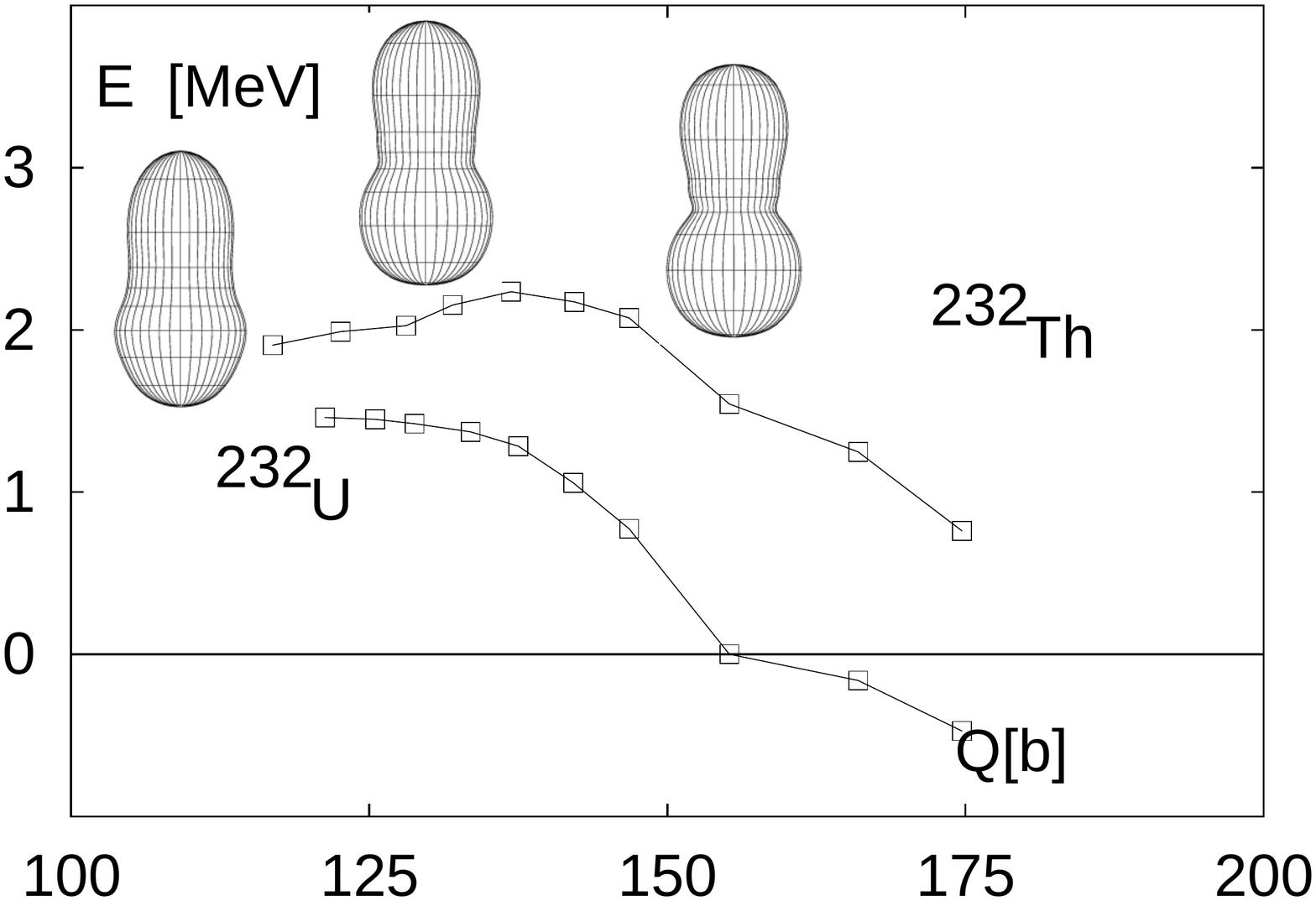}
\end{minipage}
\caption{{\protect\small Energy along a sequence of
  continuously elongating shapes parametrized by $\beta_1-\beta_8$,
 beginning at the apparent third minima with smaller octupole deformation
 $\beta_3\approx 0.3$ in $^{232}$Th and $^{234}$U.
 Quadrupole moments calculated as in Fig.1.
 Depicted shapes correspond to states in $^{232}$Th: Q=117
 (apparent IIIrd minimum), 137 and 155 b.
    }}
\end{figure}
   The results of this work may be summarized as follows:

  (i) With the shape parametrization (\ref{shape}), modifications
  of energy landscapes by including the dipole
  deformation $\beta_1$ become important beyond the IInd barrier in
   actinides.

   (ii) They remove minima with larger octpole deformations
     found in \cite{Stefan,Stefan1}. They also remove minima
   with smaller octupole deformations, perhaps, except $^{230,232}$Th, where
    only a very shallow minima can exist.
   Time-consuming, 9D calculations are required for
   a precise (up to less than 100 keV) determination of the IIIrd barriers.

   (iii)  The IIIrd barriers predicted by the Woods-Saxon model,
     if they exist at all, do not exceed a few hundreds keV which is
  in agreement with selfconsistent mean-field results with
   realistic forces.
    They do not correspond to a type suggested originally in \cite{Stefan,Stefan1}
  which served a physical interpretation of some experimental data.

   It remains to understand the observed resonances in fission probability
   and their structure. On the theory side, with the mean-field barriers
   so shallow, one should check collective (beyond mean-field) corrections.
    One could also check barriers at higher spins.

The authors would like to thank P.~Jachimowicz for
   making some test calculations, P.~G.~Thirolf, L.~Csige and D.~Habs for valuable
  explanations of experimental results, P.~Moller for communications on
   shape parametrization and M.~Bender for selfconsistent Skyrme-BCS results.
  The support of the LEA COPIGAL fund is gratefully acknowledged.

\end{document}